\documentclass[%
 reprint,
 amsmath,amssymb,
 aps,
 superscriptaddress,
]{revtex4-2}

\usepackage{graphicx}
\usepackage{dcolumn}
\usepackage{bm}
\usepackage{xcolor} 
\usepackage{upgreek}

\begin{document}

\preprint{APS/123-QED}
\title{Tuning Charge Density Wave transitions through lattice strain in NbSe$_3$}

\author{A. Gallo--Frantz}
\affiliation{Laboratoire de Physique des Solides, Université Paris-Saclay, CNRS, 91405, Orsay, France}
\author{A.A. Sinchenko}
\affiliation{Laboratoire de Physique des Solides, Université Paris-Saclay, CNRS, 91405, Orsay, France}
\affiliation{Kotelnikov Institute of Radioengineering and Electronics of RAS, 125009 Moscow, Russia}
\author{L. Ortega}
\affiliation{Laboratoire de Physique des Solides, Université Paris-Saclay, CNRS, 91405, Orsay, France}
\author{P.D. Grigoriev}
\affiliation{L.D. Landau Institute for Theoretical Physics, 142432, Chernogolovka, Russia}
\affiliation{National University of Science and Technology ``MISIS'', Moscow 119049, Russia}
\author{V.D. Kochev}
\affiliation{Kotelnikov Institute of Radioengineering and Electronics of RAS, 125009 Moscow, Russia}
\affiliation{National University of Science and Technology ``MISIS'', Moscow 119049, Russia}
\author{P. Godard}
\author{P.-O. Renault}
\affiliation{Institut Pprime, CNRS-Université de Poitiers-ENSMA, 86962, Futuroscope-Chasseneuil Cedex, France}
\author{D. Thiaudière}
\affiliation{Synchrotron SOLEIL, L'Orme des Merisiers, 91190, Saint-Aubin, France}
\author{D. Le Bolloc'h}
\affiliation{Laboratoire de Physique des Solides, Université Paris-Saclay, CNRS, 91405, Orsay, France}
\author{V.L.R Jacques}
\affiliation{Laboratoire de Physique des Solides, Université Paris-Saclay, CNRS, 91405, Orsay, France}

\date{\today}

\begin{abstract}

The Charge Density Wave (CDW) state is a perfect example of a combined structural and electronic state, both characterized by a periodic lattice distortion and an electronic modulation of condensed electrons, resulting from electron-phonon coupling. They are thus prone to be tuned by lattice strain. NbSe$_3$ is a prototypical example of CDW states, with a chain-like structure displaying two CDWs at 145K and 59K, with wavevectors along and inclined with respect to the chain axis $\textbf{b}$. Here, we report on the evolution of the lattice structure and CDW properties in the quasi-one-dimensional charge-density-wave system NbSe$_3$ under tensile stresses applied along and perpendicular to the chains' axis $\textbf{b}$ by a combination of X-ray diffraction and transport measurements. We find that the lattice structure show exotic Poisson's coefficients and strongly anisotropic Young's moduli while the evolution of the electrical resistance demonstrates shifts of both charge-density-wave critical temperatures that are strongly correlated with the lattice parameters. The two CDW transitions display different behaviours under applied stresses, and suggest that a modification of the band curvature leads to the observed transport signatures. 

\end{abstract}

\maketitle


\section{\label{sec:Introduction} Introduction}

The sensitivity of electronic states to the host lattice through electron-phonon coupling makes strain a key to study electronic phases and more generally, to explore the phase diagram of condensed matter systems. The evolution of collective electronic states such as superconductivity or charge density waves (CDW) under deformation has been at the center of a growing number of studies lately. CDWs have been drawing attention for decades due to their ubiquity in low-dimensional correlated systems \cite{monceau_electronic_2012, gruner_density_2019} and more recently, because of their competition with other electronic phases such as superconductivity  \cite{regueiro_superconductivity_1992, tranquada_evidence_1995, zocco_pressure_2015, wang_competition_2021}. Recently, the application of mechanical tensile or compressive strain in CDW materials revealed strong variations of the critical temperature in quasi-1D systems like K$_{0.3}$MoO$_{3}$ \cite{zybtsev_effect_2025}, TaS$_3$ \cite{preobrazhensky_nonlinear_1985} and NbSe$_3$ \cite{lear_stress_1984, tseng_separation_1993, kuh_nbse3_1998}  as well as in quasi-2D systems like the RTe$_{3}$ systems (R a rare-earth element) \cite{straquadine_evidence_2022, gallofrantz_charge_2024, singh_emergent_2024} and Transition Metal Dichalcogenides \cite{blundo_strain-tuning_2021}.\\

\begin{figure}[h]
    \centering
    \includegraphics[width = 0.90\linewidth]{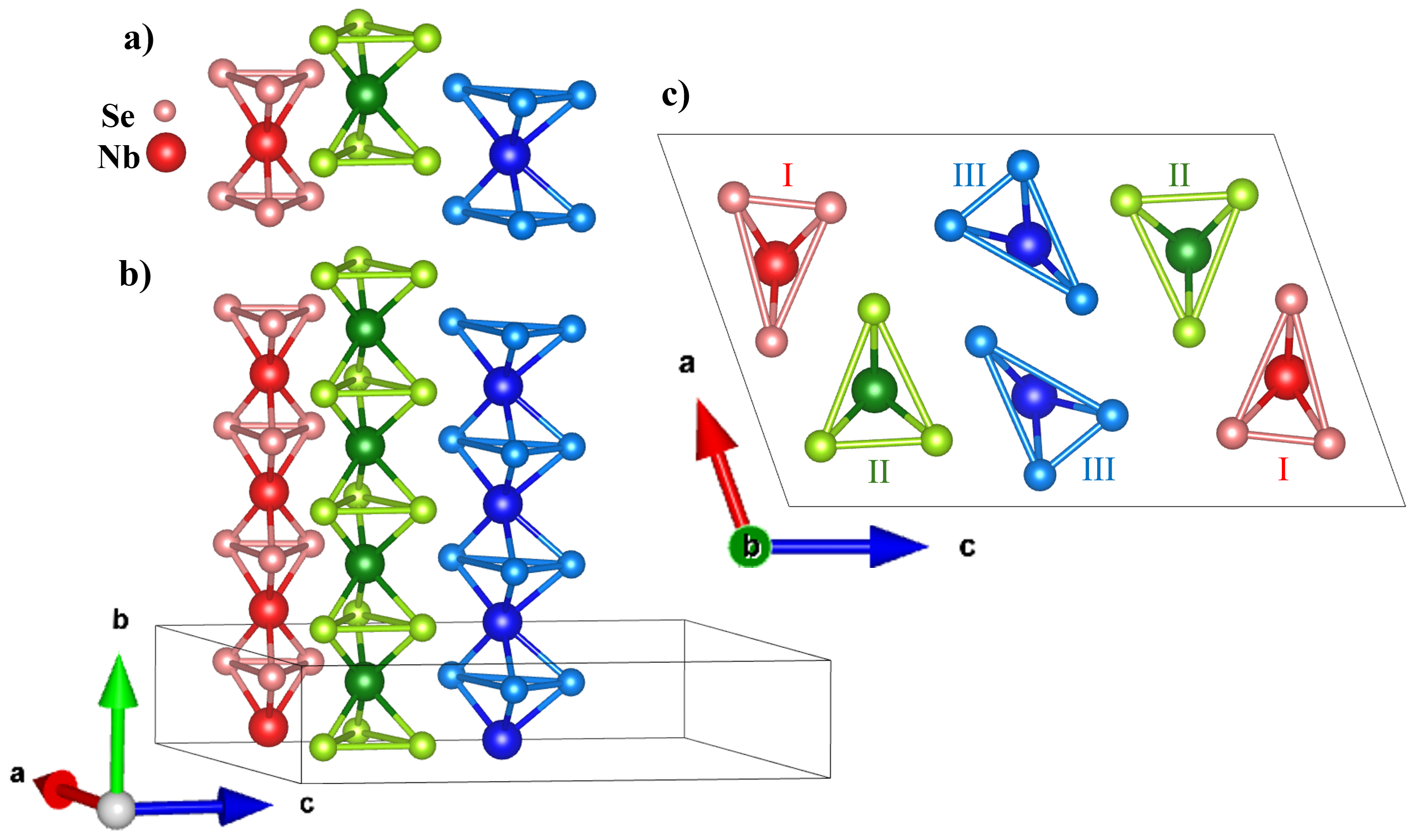}
    \caption{NbSe$_3$ crystal structure with a) the three different types of prisms for type I, II and III chains. b) 3D crystal structure made of three different and weakly coupled 1D chains along $\textbf{b}$. c) Projection of the unit cell in the $(\textbf{a}, \textbf{c})$ plane.}
    \label{fig:NbSe3_crystal_structure}
\end{figure}

\noindent NbSe$_3$ is one of the archetypes of a quasi-one-dimensional CDW system \cite{monceau_electronic_2012, gruner_density_2019}. As illustrated in Fig.\ref{fig:NbSe3_crystal_structure}, it is made of 3 types of chains resulting from a stack of NbSe$_3$ prisms with a triangular base and crystallizes in a monoclinic structure ($P2_1/m$) with lattice parameters a = 10.009 \AA, b = 3.481 \AA, c = 15.629 \AA\ and $\upbeta = 109.5^{\circ}$, where the chains are parallel to the $\textbf{b}$-axis \cite{hodeau_charge-density_1978}. Due to the favorable nesting conditions of the Fermi surface (FS), NbSe$_3$ exhibits two incommensurate charge density waves (CDW) at T\textsubscript{c1} = 145 K and T\textsubscript{c2} = 59 K with respective wavevectors $\textbf{q}_1 \approx (0, 0.243, 0)$ and $\textbf{q}_2 \approx (0.5, 0.263, 0.5)$ \cite{chaussy_phase_1976, fleming_x-ray_1978, schafer_high-temperature_2001}. These new periodicities partially gap the bands at Fermi level leading to increases of electrical resistivity below each of the two critical temperatures \cite{nicholson_dimensional_2017, nicholson_role_2020}. However, NbSe$_3$ keeps a metallic behavior, even at low temperature, because the two successive nestings are incomplete.\\

\noindent In NbSe$_3$, it was found that the application of tensile stress $\upsigma$ along $\textbf{b}$ results in a decrease of both T\textsubscript{c1} and T\textsubscript{c2} but in a different way : T\textsubscript{c1} decreases linearly with a slope $dT_{c1}/d\sigma = -4.4$ K/GPa while T\textsubscript{c2} displays less variations and a quadratic dependence on stress \cite{lear_stress_1984, tseng_separation_1993, kuh_nbse3_1998}. The decrease of T\textsubscript{c} was attributed to the transverse contraction due to Poisson's ratio \cite{lear_stress_1984}. The estimated change of T\textsubscript{c} using isothermal compressibility \cite{yamaya_pressure_1983} and Young's modulus along the chains \cite{brill_elastic_1982} gave a decent agreement for T\textsubscript{c1} but a big discrepancy for T\textsubscript{c2}. Note that this work was performed without any direct measurement of lattice parameters. Another explanation found in the literature is based on the change in fermiology in NbSe$_3$ under uniaxial deformation. This is supported by Shubnikov-de Haas measurements that display a decrease of the Fermi surface area when the deformation increases \cite{kuh_nbse3_1998}.\\

\noindent On the other hand, much less is known about the application of tensile stress along the $\textbf{c}$-axis. To our knowledge, only one work reported the influence of a $\textbf{c}$-axis deformation in NbSe$_3$ \cite{kowada_development_2007}. However, the main goal of this work was to report on the development of a uniaxial stress apparatus to elongate soft crystalline samples. NbSe$_3$ was presented as a test system for this experimental technique. The samples used in this study were imperfect so that the conclusions drawn about the physical properties of NbSe$_3$ under deformation can be debated. Thus, the exact mechanism of the CDW evolution in NbSe$_3$ under tensile stress along the $\textbf{c}$-axis required new experiments. \\

\noindent Here, we present a dedicated work to follow the CDW evolution in NbSe$_3$ under tensile stress along both $\textbf{b}$ and $\textbf{c}$ axis. To do so, we measured the lattice parameters and electronic transport properties in a large temperature range to track the evolution of the two CDW transitions while applying uniaxial stresses in these two orthogonal directions.\\

\section{\label{sec:Experimental_results} Experimental results}

\begin{figure}[h]
    \centering
    \includegraphics[width = 0.75\linewidth]{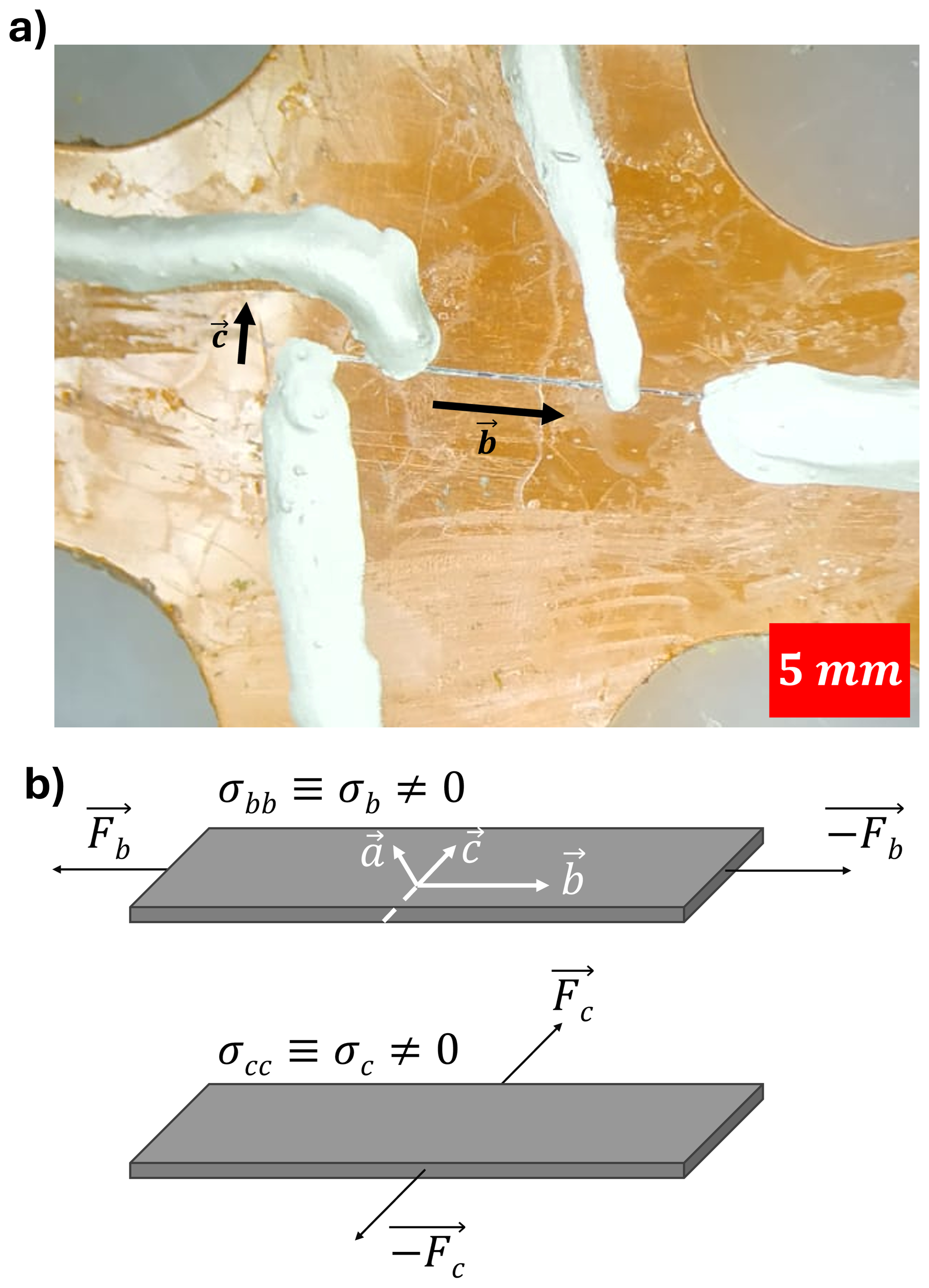}
    \caption{a) NbSe$_3$ sample (narrow metallic-grey line) glued on the cross-shaped deformable substrate with the $\textbf{b}$ and $\textbf{c}$ crystallographic axis aligned with the substrate branches. Four silver paint electrical contact are made for current injection and voltage measurement in four-bar geometry. b) Schematic drawing of NbSe$_3$ under applied stresses along \textbf{b} and \textbf{c}.}
    \label{fig:NbSe3_glued_kapton_substrate}
\end{figure}

For these experiments, we selected a high-quality NbSe$_3$ single crystal $\sim$20 $\upmu$m wide and $\sim$5 mm long. This crystal was glued at the center of a 125$\upmu $m-thick polyimide cross-shaped substrate, with $\textbf{b}$ and $\textbf{c}$ in-plane directions aligned with the arms of the polyimide cross, as shown in Fig.\ref{fig:NbSe3_glued_kapton_substrate}. The sample was then mechanically exfoliated down to 2 $\upmu$m to get homogenous deformation during mechanical deformation.


\noindent The polyimide substrate, with the sample glued on it, was then installed in the biaxial tensile stress device described in the supplementary information of Ref.\cite{gallofrantz_charge_2024}. The four branches of this cross-shaped substrate are attached to four independent motors that can pull on each branch independently. The forces applied along the arms were measured with calibrated force sensors. The center of the cross-shaped substrate lied on the cold finger of a helium-flow Konti-Micro cryostat from CryoVac GmbH, allowing to reach temperatures in the range 15-380K. All these parts are within a vacuum chamber that is close with a lid topped by a 300 $\upmu$m-thick polyether-ether-ketone (PEEK) dome to perform XRD measurements in reflection geometry.

\newpage
\subsection{Lattice evolution under uniaxial tensile stresses}

\begin{figure}[h]
    \centering
    \includegraphics[width = 1.00\linewidth]{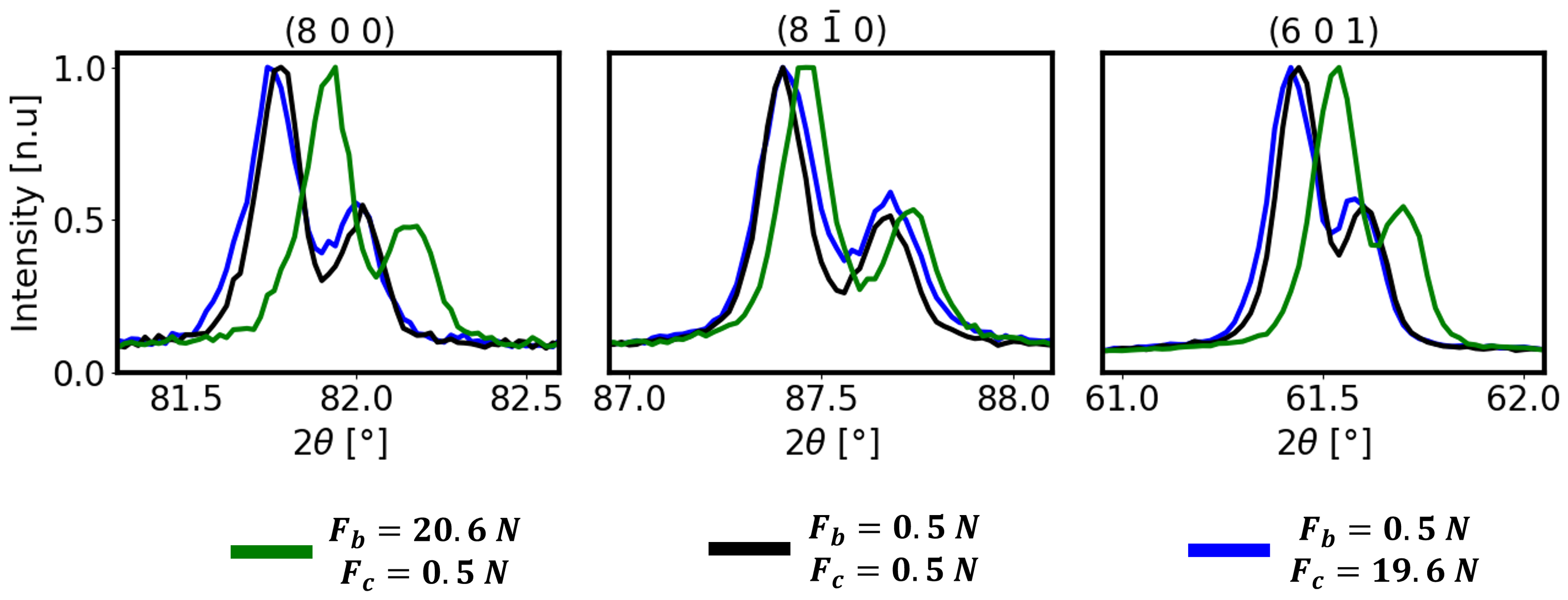}
    \caption{Normalized intensities obtained around the 800, 8$\bar{1}$0 and 601 Bragg reflections at 300 K as a function of 2$\uptheta$ for different sets of applied stresses: $\sim$20 N along $\textbf{b}$ only (green curves), $\sim$20 N along $\textbf{c}$ only (blue curves) and without stress (black curves). Each Bragg reflection displays two peaks corresponding to the two components K$_{\upalpha 1}$ and K$_{\upalpha 2}$ of the x-ray beam.}
    \label{fig:Bragg_peaks_along_tth}
\end{figure}

The lattice parameters were measured \textit{in-situ} by XRD with a 8keV beam generated by a copper rotating anode source (Rigaku RU-300B) equipped with multilayer optics that keep the K$_{\upalpha 1}$ and K$_{\upalpha 2}$ emissions lines. The sample, in the biaxial tensile stress device, was positioned at the center of rotation of a Eulerian 4-circle diffractometer and probed in reflection geometry. Diffracted beams are detected with a 2D detector located at 82 cm downstream of the sample. To get all lattice parameters, we measured the rocking curves of 3 non-collinear Bragg reflections (800, 8$\bar{1}$0 and 601). The obtained 3D volumes around these 3 positions of reciprocal space were converted into q\textsubscript{x}, q\textsubscript{y}, q\textsubscript{z} volumes and projected along the 2$\uptheta$ direction to compute all lattice parameters a, b, c, $\upalpha$, $\upbeta$ and $\upgamma$ for each set of applied forces. All lattice parameters measurements were performed at 300K, seen the weak thermal expansion of NbSe$_3$ \cite{hodeau_charge-density_1978}.\\
\noindent The projections of the measured intensities along the 2$\uptheta$ are shown in Fig.\ref{fig:Bragg_peaks_along_tth} for 3 different sets of applied forces: 0.5N on each branch (considered to be pristine state), $\sim$20 N along $\textbf{b}$ only and $\sim$20 N along $\textbf{c}$ only.



\begin{figure}[h]
    \centering
    \includegraphics[width = 1.00\linewidth]{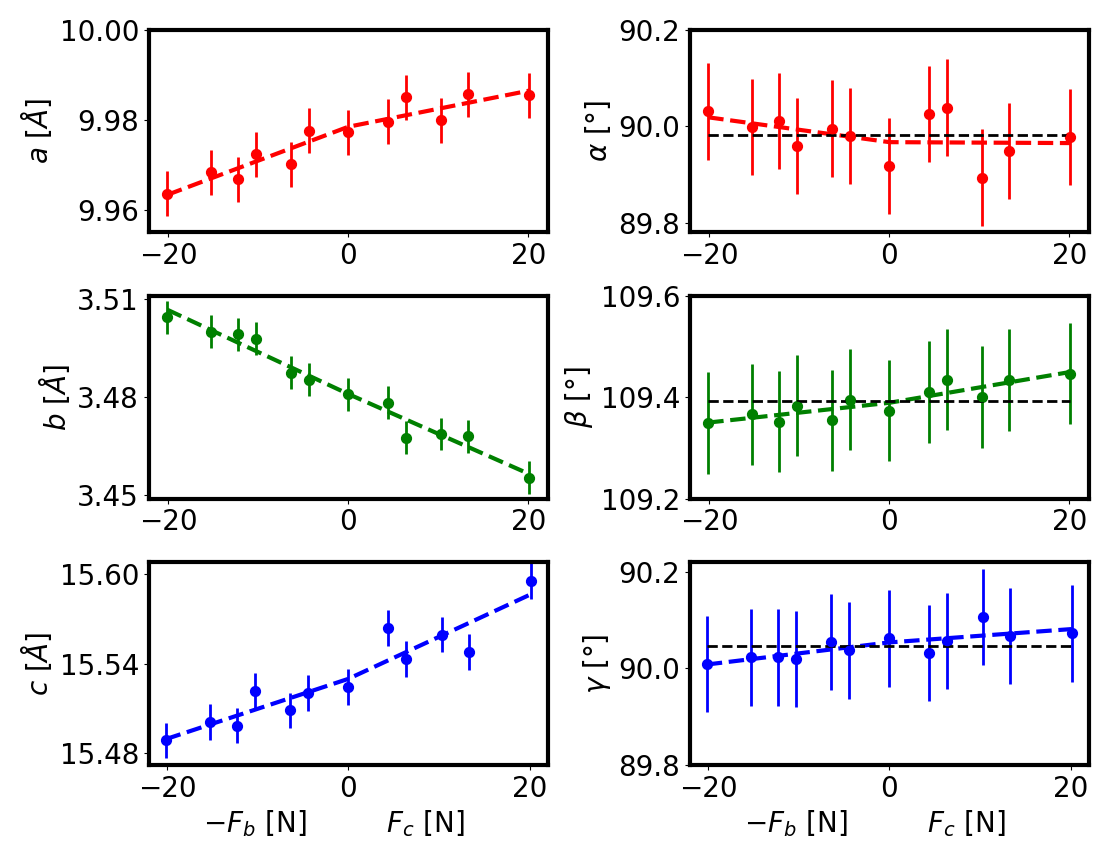}
    \caption{Evolution of the 6 lattice parameters a, b, c, $\upalpha$, $\upbeta$ and $\upgamma$ as a function -F\textsubscript{b} and F\textsubscript{c}. Two different linear functions are used to fit the curves as a function of -F\textsubscript{b} and F\textsubscript{c} (colored dashed lines) for all curves. A black dashed line corresponding to the average angles is plotted for $\upalpha$, $\upbeta$ and $\upgamma$.}
    \label{fig:lattice_parameters_as_a_function_Fb_Fc}
\end{figure}

\noindent The position of the obtained intensity distributions changes with applied stresses, especially along \textbf{b}, with no change in width, showing that the applied stresses indeed lead to lattice parameter changes, with no measurable damage. To be more precise, we computed the intensity distributions in q\textsubscript{x}, q\textsubscript{y}, q\textsubscript{z} reciprocal space coordinates, and used the vectorial position of the maxima to retrieve all 6 lattice parameters of the structure (a, b, c, $\upalpha$, $\upbeta$ and $\upgamma$), considering that the monoclinic structure would eventually not be preserved during deformation. Their evolution is shown in Fig.\ref{fig:lattice_parameters_as_a_function_Fb_Fc} as a function of -F\textsubscript{b} and F\textsubscript{c} for visual continuity reasons.

\noindent Concerning the angles $\upalpha$, $\upbeta$ and $\upgamma$ first, although they could be thought to display small linear variations, the changes are less than 0.06$^\circ$ i.e. smaller than the error bars. We thus consider the contribution of shear negligible, and that the monoclinic structure is preserved during deformation. The average values (black dotted lines in Fig.\ref{fig:lattice_parameters_as_a_function_Fb_Fc}) correspond to the ones found in the pristine state: $\upalpha=89.98\pm 0.10^\circ$; $\upbeta=109.40\pm 0.10^\circ$ and $\upgamma=90.03\pm 0.10^\circ$.

\noindent As expected from Fig.\ref{fig:Bragg_peaks_along_tth}, a, b and c are affected by a stress applied along \textbf{b} and along \textbf{c}, with linear variations in both cases, but not with the same slope. To illustrate this, the evolution of a, b and c are fitted with different linear functions for stresses applied along \textbf{b} and \textbf{c} in Fig.\ref{fig:lattice_parameters_as_a_function_Fb_Fc}. a and c clearly display two different slopes as a function of -F\textsubscript{b} and F\textsubscript{c} while b keeps the same slope.\\

\noindent From these measurements, we can get the maximum normal strains $\varepsilon_a$, $\varepsilon_b$ and $\varepsilon_c$ along a, b and c respectively, and neglect all shear strains as $\upalpha$, $\upbeta$ and $\upgamma$ are considered constant. The forces are applied along b and c, and we can relate the measured strains to the Young's moduli, Poisson's coefficients and applied stresses. In the following, $\upnu_{ij}$ is the Poisson's coefficient measured along $j$ when a stress is applied along $i$, $E_i$ is the Young's modulus along $i$ and $\upsigma_i$ is the normal stress along $i$ ($i,j$ = $a,b,c$).

\noindent When the stress is applied along $\textbf{b}$, we get the following maximum strains and can relate them to the Young's moduli, Poisson's coefficients and applied stresses as follows:
\begin{eqnarray*}
    \varepsilon_a &=& \frac{\Delta a}{a_0} = -\frac{\upnu_{ba}}{E_b}\upsigma_{b} \approx -0.15 \: \% \\\\
    \varepsilon_b &=& \frac{\Delta b}{b_0} = \frac{1}{E_b}\upsigma_{b} \approx 0.75 \:  \% \\\\
    \varepsilon_c &=& \frac{\Delta c}{c_0} = -\frac{\upnu_{bc}}{E_b}\upsigma_{b} \approx -0.25 \:  \%
\end{eqnarray*}

\noindent where $a_0$, $b_0$ and $c_0$ are the lattice parameters measured in the pristine state. In this configuration, the strains are positive along the traction direction, and negative in the orthogonal directions, which is an expected behaviour. In addition, we can compute the two Poisson's coefficients $\upnu_{ba}$ and $\upnu_{bc}$ :

\begin{eqnarray*}
    \upnu_{ba}  &=& -\frac{\varepsilon_a}{\varepsilon_b} \approx 0.20 \\\\
    \upnu_{bc} &=& -\frac{\varepsilon_c}{\varepsilon_b} \approx 0.33 
\end{eqnarray*}

\noindent which are quite usual values for solids. Similarly, when the stress is applied along $\textbf{c}$, the maximum measured strains are :
\begin{eqnarray*}
   \varepsilon_a &=& \frac{\Delta a}{a_0} = -\frac{\upnu_{ca}}{E_c}\upsigma_{c} \approx 0.10 \: \% \\\\
   \varepsilon_b &=& \frac{\Delta b}{b_0} = -\frac{\upnu_{cb}}{E_c}\upsigma_{c} \approx -0.75 \: \% \\\\
   \varepsilon_c &=& \frac{\Delta c}{c_0} = \frac{1}{E_c}\upsigma_{c} \approx 0.45 \: \%
\end{eqnarray*}

\newpage
\noindent In this case, the contraction of $b$ is much larger than the expansion of $c$ although the force is applied along c, and $\varepsilon_a > 0$, which are unexpected behaviours. The resulting Poisson's coefficients $\upnu_{ca}$ and $\upnu_{cb}$ are thus unconventional: 

\begin{eqnarray*}
    \upnu_{ca}  &=& -\frac{\varepsilon_a}{\varepsilon_c} \approx -0.22 \\\\
    \upnu_{cb} &=& -\frac{\varepsilon_b}{\varepsilon_c} \approx 1.67 
\end{eqnarray*}

\noindent Here, we have $\upnu_{ca} < 0$, meaning that $a$ expands when pulling along $\textbf{c}$, and $\upnu_{cb} > 1$, which corresponds to a stronger deformation along b than along the pulling direction. These results account for a very peculiar mechanical behaviour of NbSe$_3$ when pulling along $c$. Only a few materials, called "auxetics", have a negative Poisson's coefficient, such as 2D silica, with a predicted value $\upnu_{yx} = -0.21$ and a high piezoelectric coefficient \cite{ozcelik_stable_2014}. In addition, $\upnu_{cb}$ has a strikingly high value, well above the standard values of most materials. Only some foams exhibit Poisson's coefficients greater than one \cite{lee_anisotropic_1997}. This could be due to very different Young's moduli along the different crystallographic directions. We can get the ratio of $E_c$ and $E_b$ from the ratio of Poisson's coefficients $\upnu_{cb}/\upnu_{bc}$ :

\begin{equation*}
    \frac{E_c}{E_b}=\frac{\upnu_{cb}}{\upnu_{bc}} = \frac{1.67}{0.33}\approx 5     
\end{equation*}

\noindent As expected for an anisotropic material, $\upnu_{bc}\neq\upnu_{cb}$, and the ratio proves that the Young's moduli along \textbf{b} and \textbf{c} are highly different. The absolute value of Young's moduli has been estimated in few papers only, with strong variations from one sample to another (150 $\pm$ 100 GPa to 550 $\pm$ 280 GPa in Ref.\cite{brill_youngs_1978}). Although we do not get an absolute value of Young's moduli here either, we find that $E_c$ is 5 times larger than $E_b$, proving the high mechanical anisotropy of NbSe$_3$.

\subsection{Resistance along the chains direction under uniaxial tensile stresses}

\begin{figure}[h]
    \centering
    \includegraphics[width = 1.00\linewidth]{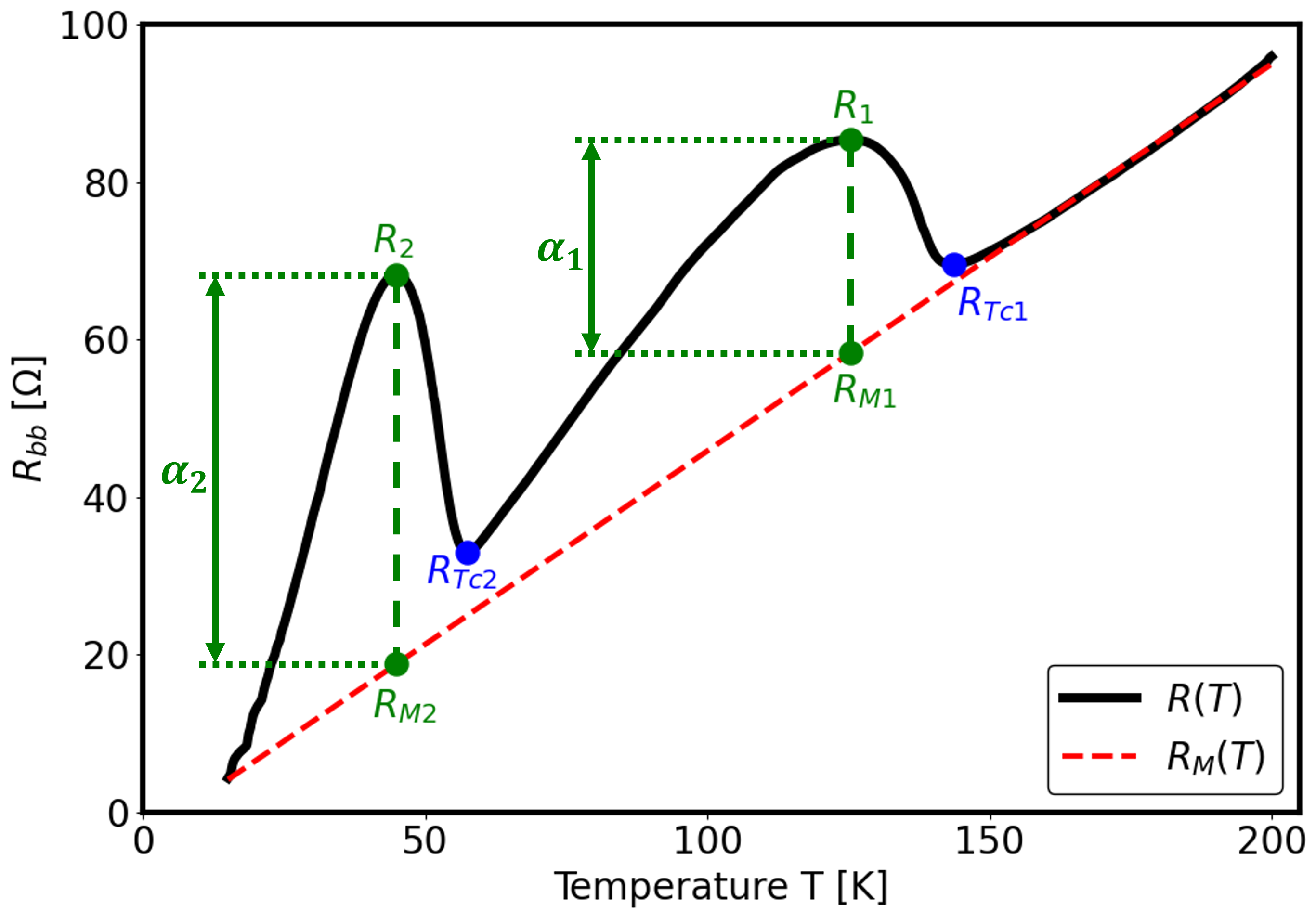}
    \caption{$R(T)$ curve measured in the pristine state (black solid line) and linear extrapolation $R_M(T)$ from the metallic part of $R(T)$ (red dotted line). The resistance values at critical temperatures $R_{Tc1}$ and $R_{Tc2}$ are marked with blue dots while the resistance increments $\alpha_1$ and $\alpha_2$ and the resistance values $R_1$, $R_2$, $R_{M1}$ and $R_{M2}$ used to compute them (see text) are marked with green dots.}
    \label{fig:NbSe3_resistance_pristine_state}
\end{figure}

To follow the transport properties of the \textit{in-situ} deformed sample, four contacts were deposited in a four-bar geometry along the chains direction ($\textbf{b}$-axis), as shown in Fig.\ref{fig:NbSe3_glued_kapton_substrate}. A Keithley 2611 Sourcemeter and a Keithley 2182a Nanovoltmeter were used to apply the current and to measure the voltage respectively. These quantities were used to compute the resistance along \textbf{b} R\textsubscript{bb}. To measure the evolution of R\textsubscript{bb} as a function of temperature and uniaxial tensile stresses, the forces F$_b$ and F$_c$ were changed at 250 K and the resistance was measured during cooling down to 15 K at a fixed rate of 1 K/min.
The R(T) curve obtained in the pristine state is shown in Fig.\ref{fig:NbSe3_resistance_pristine_state}, and corresponds to the typical R(T) behaviour as reported in the literature \cite{monceau_electric_1976, ong_conductivity_1978}. From our measurements, we get a residual-resistance-ratio $RRR = R(300 \: K) / R(4.2 \: K) =$ 100, proving the high-quality of this sample.\\

\noindent Both critical temperatures T$_{c1}$ and T$_{c2}$ are extracted from the local minima of the R(T) curves (see R$_{Tc1}$ and R$_{Tc2}$ in Fig.\ref{fig:NbSe3_resistance_pristine_state}) while resistance increments $\alpha_1$ and $\alpha_2$, corresponding to the resistance jumps below T$_{c1}$ and T$_{c2}$ respectively, are defined as follows :

\begin{equation}
    \alpha_i = \frac{R_i - R_{M,i}}{R_i}  \hspace{0.5cm},\hspace{0.5cm} i=1,2 
    \tag{II.B.1}
    \label{eq:resistance_increments}
\end{equation}\\

\noindent where R$_i$ is the resistance at the local maximum of R(T) below each CDW transition, R$_{M,i}$ is the value of R$_M(T)$ at the same temperature as R$_i$ ($i=1,2$), and R$_M(T)$ is the linear curve extrapolated in the full temperature range from the metallic part of R(T) (above T$_{c1}$) \cite{ido_pressure_1990, zhao_pressure-modulated_2026}. $R(T)$ and $R_M(T)$ are shown in Fig.\ref{fig:NbSe3_resistance_pristine_state} as black solid line and red dotted line respectively,  $R_{Tc1}$ and $R_{Tc2}$ are shown with blue dots and $R_i$, $R_{Mi}$ and $\alpha_i$ ($i=1,2$) are shown in green. 

\begin{figure}[h]
    \centering
    \includegraphics[width = 1.00\linewidth]{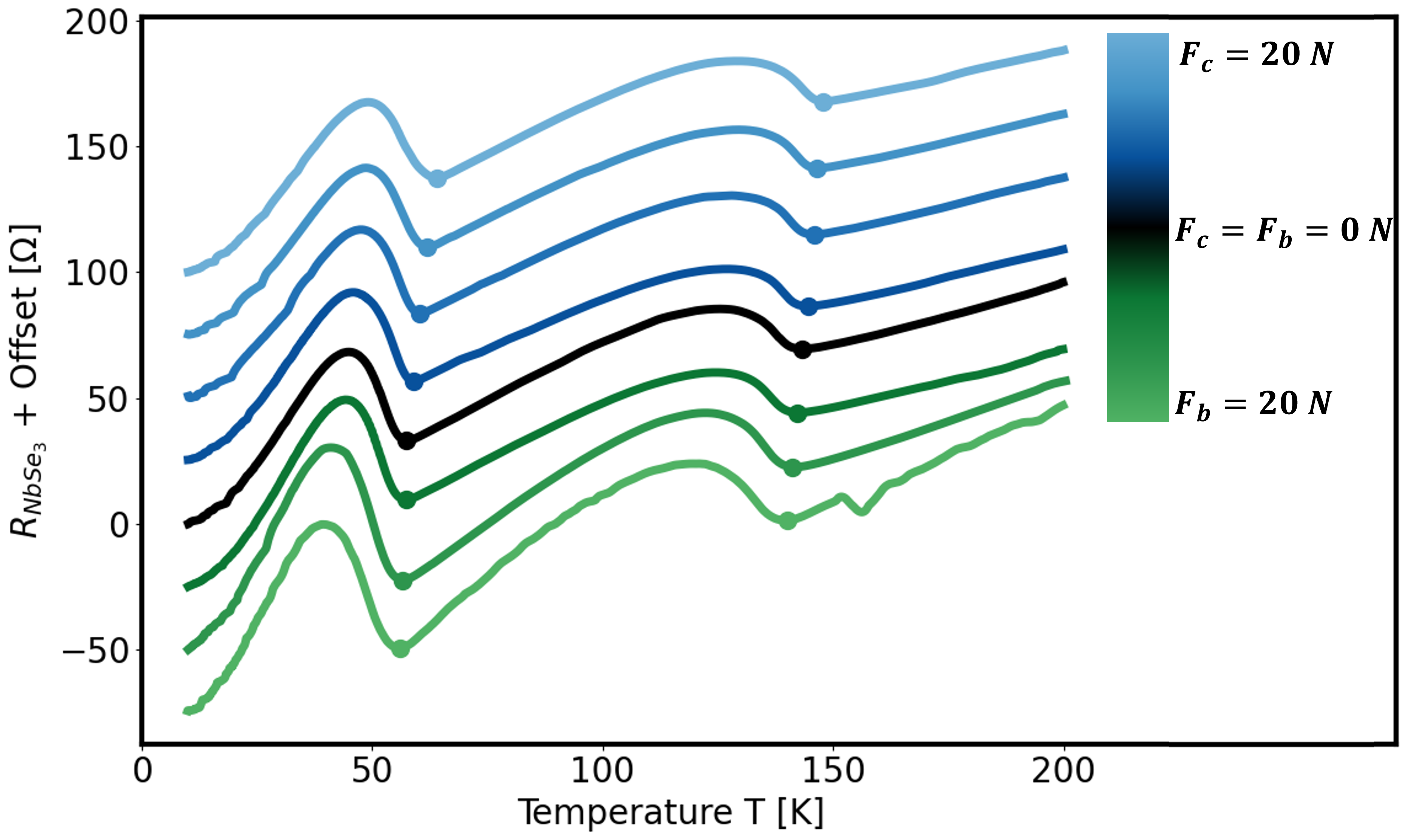}
    \caption{R(T) curves measured under tensile stress along $\textbf{b}$ (green curves), along $\textbf{c}$ (blue curves) and in the pristine state (black curve). A 25$\Omega$ offset has been added between each curve for clarity. For all curves, T$_{c1}$ and T$_{c2}$ are marked with colour dots.}
    \label{fig:NbSe3_resistance_tensile_stress}
\end{figure}

\begin{figure}[h]
    \centering
    \includegraphics[width = 1.00\linewidth]{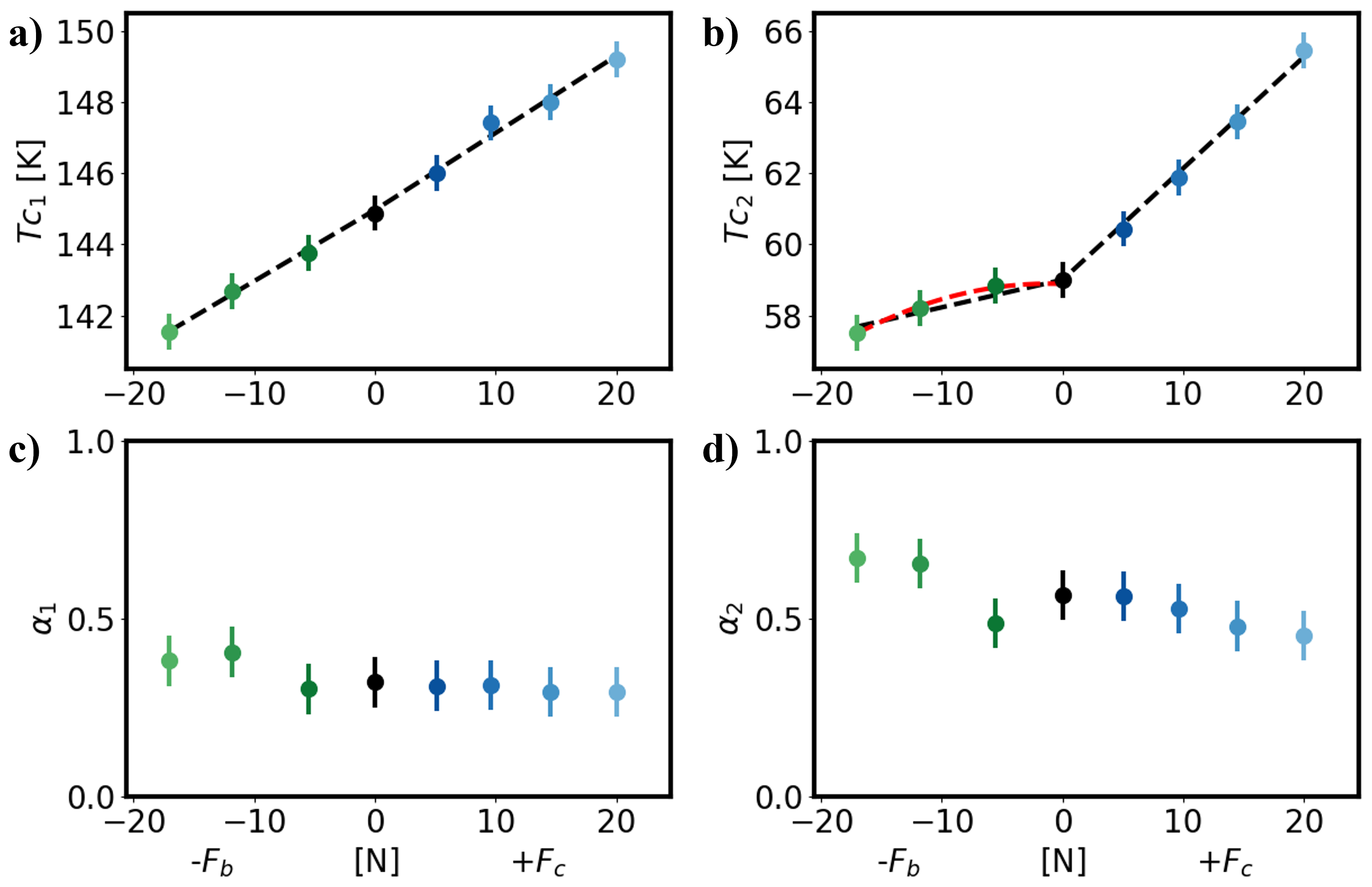}
    \caption{Evolution of a) T$_{c1}$, b) T$_{c2}$,  c) $\alpha_1$ and d) $\alpha_2$ as a function of $-F_b$ and $F_c$. For the variations of critical temperatures, the linear fits (black dashed lines) and quadratic fit (red dashed line), according to \cite{lear_stress_1984}, are shown to guide the eyes.}
    \label{fig:Tc_alpha_Fb_Fc}
\end{figure}

\noindent The R(T) curves measured under the applications of tensile forces up to 20N along \textbf{b} and \textbf{c} are shown in Fig.\ref{fig:NbSe3_resistance_tensile_stress}. The general shape of the $R(T)$ curves measured under stress is kept, the signature of the two CDW transitions appearing as a local minimum before an increase of the resistance below T$_{c1}$ and T$_{c2}$. However, the values of T$_{c1}$ and T$_{c2}$ extracted from these curves are not preserved, and seem to display continuous variations as a function of $-F_b$ and $F_c$, as shown in Fig.\ref{fig:Tc_alpha_Fb_Fc}.

\newpage
\noindent Both critical temperatures, illustrated in Fig.\ref{fig:Tc_alpha_Fb_Fc} a) and b), show a common general behavior as a function of applied stress. When the stress is applied along the chains ($\textbf{b}$-axis), the critical temperatures decrease while they increase when the stress is applied perpendicularly ($\textbf{c}$-axis). The linear variations of T$_{c1}$ as a function of stress are in line with the ones found in the literature \cite{lear_stress_1984, tseng_separation_1993, kuh_nbse3_1998, kowada_development_2007}.\\

\noindent Contrary to the results shown in Ref.\cite{kowada_development_2007}, the stress applied along \textbf{c} induces great changes on T$_{c2}$, with an increase of more than $10\: \%$, as illustrated in Fig.\ref{fig:Tc_alpha_Fb_Fc} a). In comparison, the relative variation of T$_{c1}$ is three times as small. The much smaller changes on T$_{c2}$ reported in Ref.\cite{kowada_development_2007} could be explained by the poorer sample quality and by a non-perfect deformation transfer. Indeed, in this reference, the resistance increment value for the second transition $\alpha_2$ is between two and three times less than expected for high-purity samples and the deformation method could lead to a poorer deformation transfer to the sample.\\

\begin{figure}[h]
    \centering
    \includegraphics[width = 1.00\linewidth]{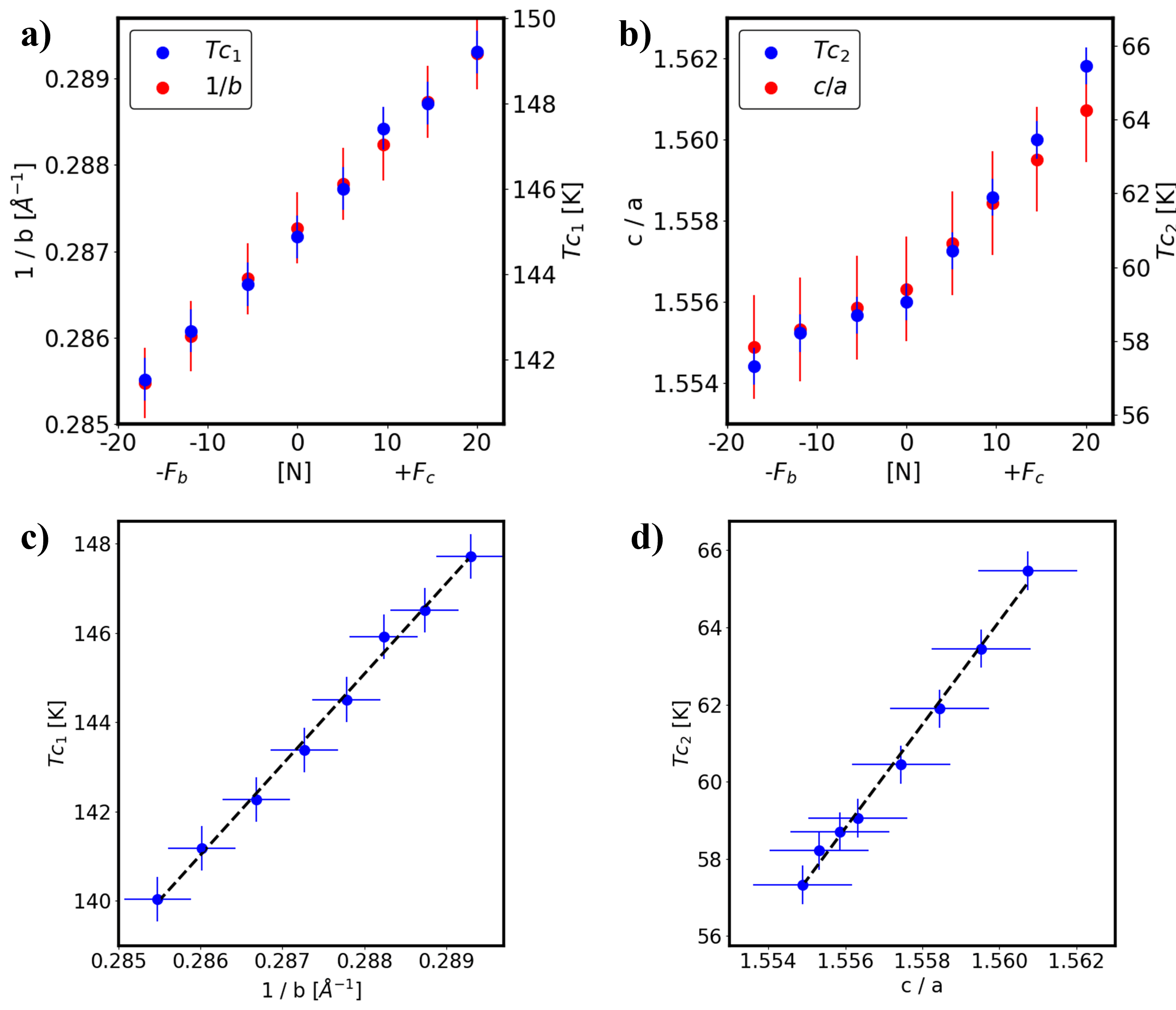}
    \caption{a) $1/b$ (red dots) and T$_{c1}$ variations as a function of the applied force along $\textbf{b}$- and $\textbf{c}$-axis. b) $c/a$ ratio (red dots) and T$_{c2}$ (blue dots) variations as a function of the applied force along $\textbf{b}$- and $\textbf{c}$-axis. c) T$_{c1}$ as a function of $1/b$. d) T$_{c2}$ as a function of $c/a$ ratio. For c) and d), regarding the error bars, the link between critical temperatures and lattice structure is linear.}
    \label{fig:correlation_between_Tc_and_lattice}
\end{figure}

\noindent However, the variations of both critical temperatures are different. T$_{c1}$ displays a single linear slope as a function of $-F_b$ and $F_c$, while T$_{c2}$ displays two distinct slopes. To go further, we can analyze the correlations between critical temperatures and lattice structure. We found in particular that T$_{c2}$ and $c/a$ (resp. T$_{c1}$ and $1/b$) behave similarly as a function of applied forces along $\textbf{b}$ and $\textbf{c}$ (see Fig.\ref{fig:correlation_between_Tc_and_lattice} a) and b)).

\noindent This connection becomes even clearer when plotting T$_{c2}(c/a)$ and T$_{c1}(1/b)$ which both show a perfect linear coupling (see Fig.\ref{fig:correlation_between_Tc_and_lattice} c) and d)). This striking feature illustrates the direct link between the lattice structure, the CDW structure and the associated transition temperatures. {\color{black} To understand this link we analyze the stress dependence of Lindhard electronic susceptibility 
	\begin{gather}
		\chi(T,\mathbf{Q})=
		-\int\frac{d\mathbf{k}}{(2\pi)^2}\frac{n_{\rm F}(\mathbf{k}-\mathbf{Q},T)-n_{\rm F}(\mathbf{k}, T)}{\epsilon(\mathbf{k}-\mathbf{Q})-\epsilon(\mathbf{k})},
		\label{Lindhard}
        \tag{II.B.2}
	\end{gather}
which determines the CDW transition temperature T$_c$ according to the following condition (Eq.\ref{eq:Tc})
\begin{equation} 
	\begin{aligned}
		U(\bm{Q}).\chi(T_c,\bm{Q})=1
	\end{aligned}
	\label{eq:Tc}
    \tag{II.B.3}
\end{equation}
where  $n_{\rm F}(\mathbf{k}, T)$ is the Fermi distribution function, $\epsilon(\mathbf{k})$ is electron energy at quasi-momentum $\mathbf{k}$, and U($\bm{Q}$) is the static combined electron-electron interaction which includes both Coulomb and phonon-mediated interaction.

\newpage
\noindent The higher the Lindhard susceptibility, the higher the transition temperature. In quasi-1D metals like NbSe$_3$, the electronic dispersion can be written as follows :
 
\begin{gather}
	\epsilon(\mathbf{k}) \approx - 2t_a \cos ({a.k_x}) - 2t_b \cos ({b.k_y}) - 2t_c \cos ({c.k_c})
    \label{spectrum3D}
    \tag{II.B.4}
\end{gather}
where $t_b \gg t_a, t_c$ for each band, in addition to the nesting quality, the Lindhard susceptibility is inversely proportional to the electron velocity $v_F$ at Fermi level along the chain direction:
\begin{equation} 
	\begin{aligned}
\chi (T,\mathbf{Q})\propto 1/v_F\propto 1/(t_b.b)	.
	\end{aligned}
	\label{chi1}
    \tag{II.B.5}
\end{equation}
The tensile strain along the chain direction has two opposite effects on electron velocity $v_F$: it increases $b$ but decrease $t_b$. {\color{black} T$_{c1}$ is thus expected to change in the same direction as observed experimentally as a function of stress (see Figs. \ref{fig:Tc_alpha_Fb_Fc}a or \ref{fig:correlation_between_Tc_and_lattice}a,c) taking into account that the bandwidth $4t_b$ changes more slowly than the lattice constant $b$ (see Suppl. Mat.).}\\

\noindent The evolution of T$_{c2}$ with stress is more complicated, because CDW$_2$ gets formed due to the remaining ungapped FS pockets. The nesting quality, which determines the size of these pockets and $\chi (T,\mathbf{Q})$, depends strongly on the bandwidths $4t_a$ and $4t_c$ in two perpendicular directions. Hence, the strain along perpendicular-to-chain $a$ and $c$ axes plays important role for T$_{c2}$.   
Our more rigorous calculations both of Lindhard susceptibility (Eq.\ref{Lindhard}) and of electron band structure evolution under stress using DFT support these simple considerations.} However, the reason why it explicitly depends on $c/a$ is not straightforward and still has to be explained.\\

\begin{figure}[h]
    \centering
    \includegraphics[width = 1.00\linewidth]{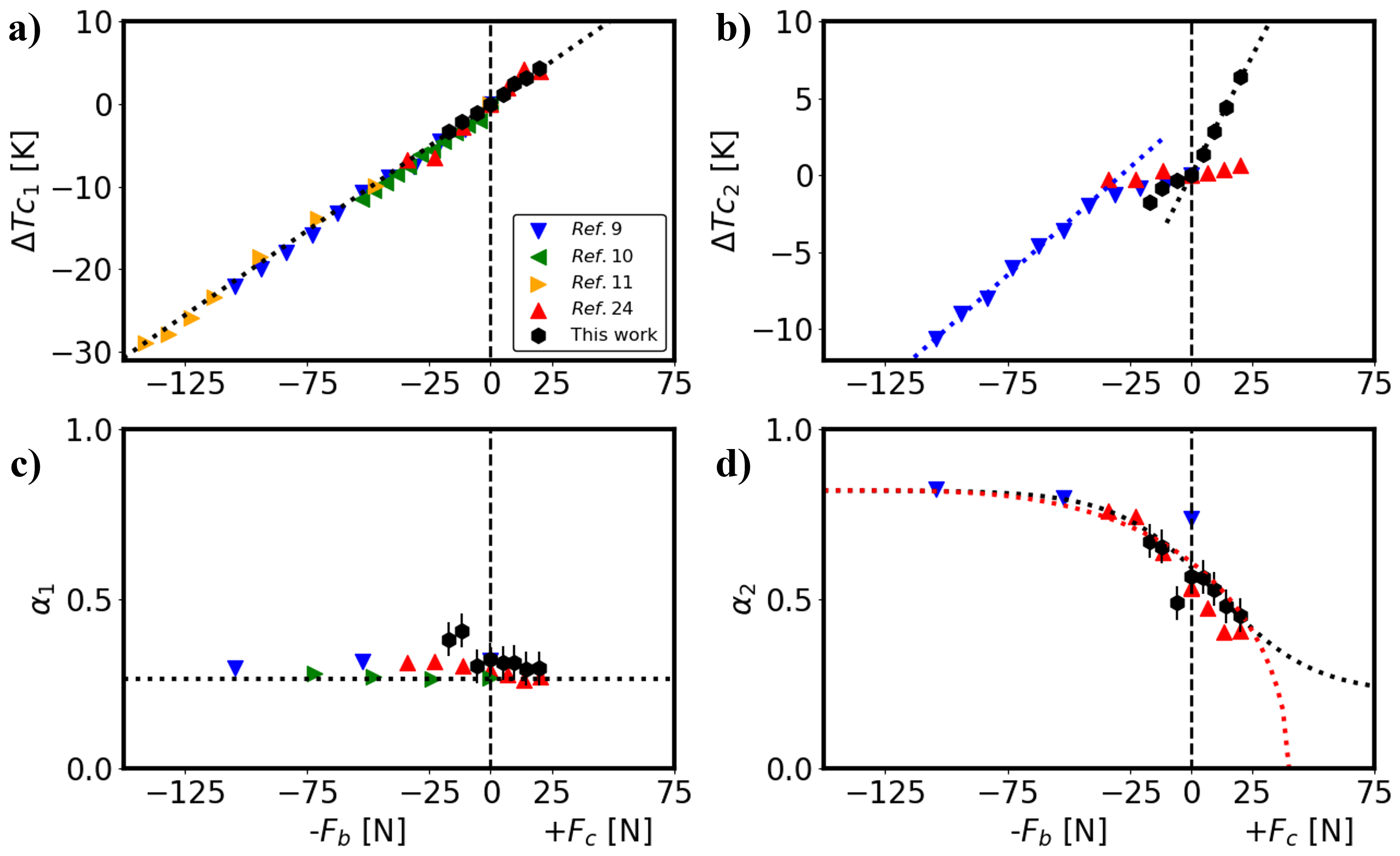}
    \caption{Evolution of a) high and b) low relative critical temperatures $\Delta T_{c1}$ and $\Delta T_{c2}$ as a function of $-F_b$ and $F_c$ respectively. Linear fits are shown with dotted lines. Evolution of c) high and d) low temperature resistance increments $\alpha_1$ and $\alpha_2$ as a function of  $-F_b$ and $F_c$. Saturation and constant regimes are shown with black dotted lines and dome in red dotted line. The measurements performed in the present study (black dots) (see Fig.\ref{fig:Tc_alpha_Fb_Fc}), are compared to values extracted from the literature (blue down triangles \cite{lear_stress_1984}, green left triangles \cite{tseng_separation_1993}, yellow right triangles \cite{kuh_nbse3_1998}, red up triangles \cite{kowada_development_2007}). Vertical dashed lines mark the pristine state. From the literature and using the correlation between T$_{c1}$ and $1/b$ (see Fig.\ref{fig:correlation_between_Tc_and_lattice}), the maximum deformation along $b$ is estimated at $\Delta b/b \approx 5.3 \:\%$}
    \label{fig:correlation_with_other_datas}
\end{figure}

\noindent Let's now analyze the evolution of $\alpha_1$ and $\alpha_2$, plotted in Fig. \ref{fig:Tc_alpha_Fb_Fc}.c) and d) as a function of $-F_b$ and $F_c$. These two quantities display around 20\% variations, between 0.5 and 0.7 for $\alpha_2$ and 0.25 and 0.45 for $\alpha_1$, with a global decrease (resp. increase) for forces applied along $\textbf{c}$ (resp. $\textbf{b}$). Our results can be compared to the values reported in the literature (\cite{lear_stress_1984, tseng_separation_1993, kuh_nbse3_1998, kowada_development_2007}, provided the applied stresses can be compared from one study to another. We did this by noticing that in those studies, like in ours, the variations of T$_{c1}$ are always linear with the deformation, so we can adjust the different values of tensile stresses to have all identical slopes $d{T_{c1}}/dF$ (see Fig.\ref{fig:correlation_with_other_datas} b). In addition, due to the different methods used to estimate absolute critical temperatures in the literature, only relative variations critical temperatures $\Delta T_c$ are shown. The same renormalized tensile stresses have been used to plot $\Delta T_{c2}$, $\alpha_1$ and $\alpha_2$ as a function of $-F_b$ and $F_c$ (see Fig.\ref{fig:correlation_with_other_datas}).

\newpage

\noindent For all references and samples, the properties of $CDW_1$ show monotonous variations : T$_{c1}$ changes linearly with the tensile stress while $\alpha_1$ seems constant. The linear decrease of T$_{c1}$ is also reported in electrical transport measurements of NbSe$_3$ done under small hydrostatic pressure ($\approx 0.5 \: GPa = 5 \:kbars$) \cite{chaussy_phase_1976, ido_pressure_1990}. However, the experiments done under higher pressure show domes in the phase diagrams with strong decreases of T$_{c1}$ and T$_{c2}$ at high pressure due to the collapse of the CDW nesting conditions \cite{regueiro_superconductivity_1992}. The linear decreases of the critical temperatures, and especially the quadratic behavior of T$_{c2}$ reported in Ref.\cite{lear_stress_1984} can be seen as beginnings of domes.\\

\noindent Regarding $\alpha$ variations, since NbSe$_3$ is partially gapped in the CDW state, it never becomes fully insulating and continues to exhibit a metallic behavior, even at low temperature. So, in the metallic state and in the CDW state, the conductivity can be written as follows :

\begin{equation}
    \sigma = \frac{n_{e, F}.e^2.\tau}{m^{\star}}
    \tag{II.B.6}
    \label{eq:conductivity}
\end{equation}\\

\noindent where $n_{e, F}$, $\tau$ and $m^{\star}$ are the density of states at Fermi level, the relaxation time and the effective mass of conduction electrons, respectively. The CDW formation is expected to modify the density of states at Fermi level as well as the effective mass of the carriers.

\newpage
\noindent If $\tau$ is not affected by the appearance of the CDW \cite{ido_pressure_1990}, the resistance increments can be rewritten as :

\begin{equation}
    \begin{split}
        \alpha = \frac{\sigma_M - \sigma}{\sigma_M}
        &= 1 - \frac{n_{CDW}}{m^{\star}_{CDW}}.\frac{m^{\star}_{M}}{n_{M}} \\
        &= 1-\frac{n_M - \Delta n}{n_M}.\frac{m^{\star}_{M}}{m^{\star}_{CDW}} \\\\
        & \Rightarrow 1-\alpha = \frac{n_M - \Delta n}{n_M}.\frac{m^{\star}_{M}}{m^{\star}_{CDW}}
    \end{split}
    \tag{II.B.7}
    \label{eq:resistance_increments_2}
\end{equation}\\

\noindent where $n_M$ and $n_{CDW}$ are the density of states in the metallic and CDW states respectively and $\Delta n$ denotes the reduction of the density of states at Fermi level due to the CDW formation. $m^{\star}_M$ and $m^{\star}_{CDW}$ are the effective masses of charge carriers in the metallic state and in the CDW state respectively.

\noindent {\color{black} The changes of $n_{e, F}$, $\tau$ and $m^{\star}$ caused by CDW  with imperfect nesting in quasi-1D metals were theoretically studied recently \cite{tsvetkova_resistivity_2025}. Although the Fermi surface (FS), the electron dispersion and the effective mass in a CDW state may change strongly, $n_{e, F}$ and $\tau \propto 1/n_{e, F}$ do not change considerably even if the large part of the FS becomes gapped by CDW \cite{tsvetkova_resistivity_2025}. This happens because the remaining ungapped electron states modify their dispersion, which  increases their $m^{\star}$ and $n_{e, F}$ and compensates the leave of gapped states. Hence, the assumption of small changes of $\tau$ due to the CDW formation is justified.}

\noindent The small variations of $\alpha_1$ could be due to relatively small changes of effective mass or density of states participating to the formation of CDW$_1$ in this range of tensile stresses or that some part of the conductivity compensates others. 

\noindent Moreover, the dome-shaped decrease of $\alpha_1$ is reported in NbSe$_3$ under pressure \cite{yasuzuka_coexistence_2005} and an increase is reported in Ta-doped NbSe$_3$ \cite{kawabata_impurity_1985}. In both cases, the CDW orders are suppressed with a large decrease of the conductivity in the metallic state for Ta-doped NbSe$_3$. This highlight that the low-temperature CDW is more sensitive than the high-temperature one to structural changes and introduction of defects.\\

\noindent On the other hand, the properties of CDW$_2$ display more complex behaviors : rather than showing two different slopes when pulling along $\textbf{b}$- and $\textbf{c}$-axis, T$_{c2}$ displays linear variations with large changes for high stresses applied along $\textbf{b}$- and $\textbf{c}$-axis but smaller variations for moderate forces along $\textbf{b}$. The global variation could be considered cubic with an inflection point lying in the area of small deformations along $\textbf{b}$-axis. Concerning $\alpha_2$, the data collection shows a saturating behaviour (black dotted line) or a dome-like shape (red dotted line) with a linear variation in the area of small stresses. Such type of variations have been observed under hydrostatic pressure \cite{ido_pressure_1990}.\\

\newpage
\noindent The increase of $\alpha_2$ under tensile stress along $\textbf{b}$-axis can suggest an increase of $\Delta n$ which means that the part of the density of states participating to the CDW increases, while the critical temperature T$_{c2}$ decreases. But it can also suggest an increase of the effective mass tensor component $m^\star_{CDW}$ in the CDW state, so a decrease of the band curvature in the $k_y$ direction due to the gap opening. On the other hand, the decrease of $\alpha_2$ under tensile stress along $\textbf{c}$-axis suggest the opposite effects (see Eq.\ref{eq:resistance_increments_2}).\\

\begin{figure}[h]
    \centering
    \includegraphics[width = 1.00\linewidth]{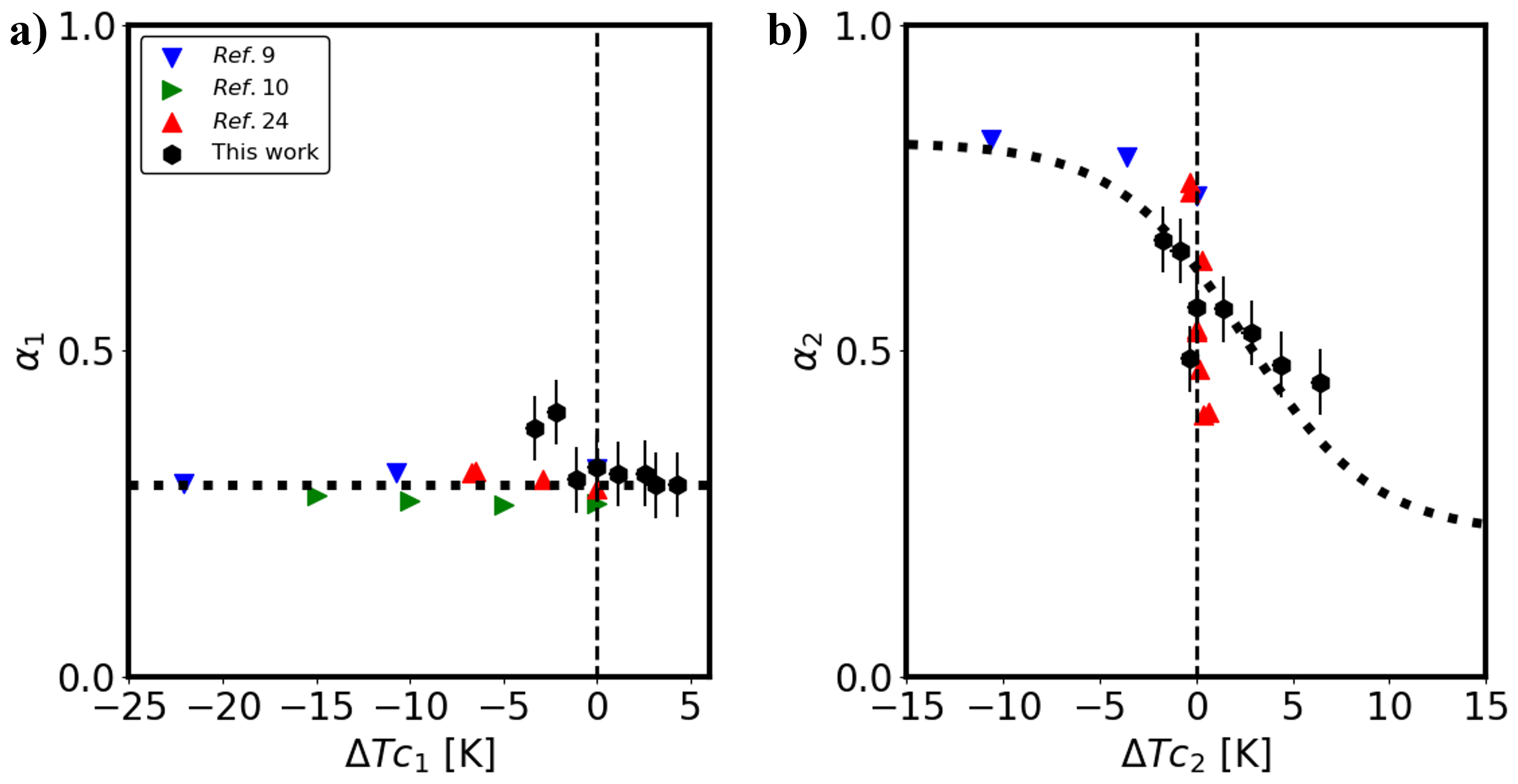}
    \caption{Evolution of a) $\alpha_1$ as a function $\Delta T_{c1}$ and b) $\alpha_2$ as a function $\Delta T_{c2}$. In both panels, black vertical dashed lines show the pristine state, without deformation, and fits are shown with black dotted lines. For the high-temperature $CDW_1$ the data show a constant regime while it seems to saturate for the low-temperature $CDW_2$. The measurements have been done on two samples : sample 1 (black dots) (see Fig.\ref{fig:Tc_alpha_Fb_Fc}), and some values were extracted from the literature (blue down triangles \cite{lear_stress_1984}, green left triangles \cite{tseng_separation_1993}, red up triangles \cite{kowada_development_2007}).}
    \label{fig:test_BCS}
\end{figure}

\noindent Finally, in quasi-1D compounds, it is reasonable to assume that the resistance increments $\alpha$ is directly proportional to the loss of electronic states that participate to the conduction and then to the gap $|\Delta|$. It is then possible to follow the evolution of $|\Delta|$ through $\alpha$ as a function of $\Delta T_c$ for both CDW transitions. This is depicted in Fig.\ref{fig:test_BCS}.

\noindent In Peierls theory, the BCS relation exhibits the linear relation between $|\Delta|$ and T$_c$. This relationship, illustrated for both CDWs under tensile stress in Fig.\ref{fig:test_BCS}, is not linear, and thus does not follow Peierl's theory : for the low-temperature CDW, the data exhibit a saturation regime with a linear decrease for small deformations while the behavior for the high-temperature CDW is constant. So, if $\alpha$ is directly proportional to $|\Delta|$, none of the two CDWs in NbSe$_3$ under stress follow the BCS equation for the gap.\\

\noindent However, if $\alpha$ is not directly linked to the gap, it reflects changes of the conductivity in the CDW state, since the metallic behavior of NbSe$_3$ seems to be unchanged by the application of tensile stress. These conductivity changes in the CDW state, should thus be attributed to variations of the density of states and of band curvature variations, modifying the nesting conditions.\\

\noindent In this work, we measured both the evolution of the lattice structure and transport properties through the two CDW transitions, as a function of uniaxial tensile stresses applied along and perpendicular to the chains' direction $\textbf{b}$. We found unusual Young's modulus $\upnu_{cb}$ for tensile stress applied along c-axis as well as direct relationship between the structural changes and transition temperatures of the two CDWs, with a clear coupling of T$_{c_1}$ (resp. T$_{c_2}$) with the $b$ lattice parameter (resp. $c/a$ ratio). In addition, the transition temperature changes T$_{c_1}$ and T$_{c_2}$ and resistivity jumps at the two transitions $\alpha_1$ and $\alpha_2$ are in line with the other works found in the literature. The link between $\alpha$ and T$_c$ does not follow the BCS relation and the changes of $\alpha$ under tensile stresses should be attributed to a modification of the band curvature modifying the nesting conditions.

\begin{acknowledgments}
This work was supported by ANR-RSF Grant no. ANR-21-CE30-0055 “BISCEPS-QM". We acknowledge Synchrotron SOLEIL for providing beamtime.
\end{acknowledgments}

\newpage
\nocite{*}
\newpage
\bibliography{NbSe3_tensile_stress_structure_Tc.bib}
\end{document}